\documentclass[runningheads]{llncs}
\usepackage[T1]{fontenc}
\usepackage{xcolor}
\usepackage{amsmath}
\usepackage{booktabs}
\usepackage{tabularx}

\usepackage{orcidlink}

\usepackage{graphicx}
\begin{document}
\title{The Hidden Cost of Contextual Sycophancy: an AI Literacy Intervention in Human–AI Collaboration
}
\titlerunning{The Hidden Cost of Contextual Sycophancy}
%
\author{Cansu Koyuturk\inst{1} \orcidlink{0009-0005-7562-7400} \and
Sabrina Guidotti\inst{1} \orcidlink{0009-0005-8606-3276} \and
Dimitri Ognibene\inst{1} \orcidlink{0000-0002-9454-680X} }
\authorrunning{C. Koyuturk et al.}
%
\institute{
Universit\`a degli Studi di Milano-Bicocca, Italy\\
\email{\{c.koyutuerk, s.guidotti2\}@campus.unimib.it}\\
\email{dimitri.ognibene@unimib.it}
}

\maketitle              
\begin{abstract}
Large Language Models (LLMs) are increasingly used in educational settings as interactive tools for collaboration. However, their tendency toward sycophancy, aligning with user beliefs even when incorrect, raises concerns for learning and decision-making, especially for less knowledgeable users. This study investigates how sycophantic alignment emerges in authentic multi-turn human–AI interactions and whether interventions targeting increasing AI literacy and prompting competencies can mitigate its effects. In a controlled mixed-design experiment, 60 participants completed analytical survival ranking tasks by first generating individual rankings and then making final decisions after collaborating with an AI assistant, both before and after receiving either general or sycophancy-focused prompting training.
Preliminary results show that LLMs are highly sensitive to user input: lower-quality initial responses lead to poorer AI advice, suggesting that the model mirrors or incorporates user reasoning rather than correcting it or offering better alternatives that are missing or less frequent in the conversation. Critically, the propagation of user errors into AI responses significantly reduced both the quality of AI feedback and final user task performance, revealing a form of contextual sycophantic dependence. While the intervention did not eliminate the propagation of contextual errors, it significantly improved AI advice by reducing the direct mirroring of incorrect user rankings.
These findings suggest that prompting and AI literacy alone may be insufficient to ensure epistemically independent AI support, highlighting the need for system-level approaches that better promote critical engagement in human–AI collaboration.

\keywords{Human--AI collaboration \and LLM sycophancy \and Prompting \and AI literacy \and Dependence \and Decision-making \and Problem-solving}
\end{abstract}
\section{Introduction}
Although commercial Large Language Models (LLMs) are not designed with educational goals in mind, they are increasingly adopted in higher education as interactive tools, with their effectiveness largely depending on how they are pedagogically structured and integrated into learning activities \cite{vendrell2026scaffolding,koyuturk2023developing}. LLMs can act as collaborative partners that actively contribute to idea generation, critique, and reasoning in learning contexts \cite{yan2025agentic}. However, their benefits depend on how models handle multi-turn exchanges, particularly when users hold incomplete or incorrect beliefs, a condition especially common among learners. 
One emerging concern is \textit{sycophancy}, the tendency of LLMs to align with user beliefs or preferences, even when those beliefs are incorrect \cite{sharma_towards_2024,doi:10.1126/science.aec8352}. Such behavior might validate user presumptions or biased views, resulting in misinformation or poor advice \cite{o2026few}. Rather than correcting misconceptions, LLMs may reinforce them, leading to poorer reasoning outcomes and increased reliance on flawed guidance \cite{bo2026invisible}. This is particularly concerning in contexts where users lack expertise, as AI feedback can be influenced by users’ own prior inputs \cite{arvin2025check} and raises a broader question about the epistemic role that LLMs play in learning. LLMs are often positioned as supportive collaborators in learning, resembling a more knowledgeable guide \cite{vygotsky1978mind}. However, when models align with user beliefs rather than task-relevant accuracy, they may fail to provide corrective scaffolding \cite{van2010scaffolding}. In such cases, the AI may function not as a better-informed partner, but as a feedback mechanism that reproduces the learner’s misconceptions, without introducing alternative perspectives or expanding understanding \cite{deng2025deconstructing}. We refer to this dynamic as \textit{sycophantic dependence}: a collaboration pattern in which users’ initial errors shape AI feedback, which in turn reinforces those same errors in users’ final decisions. 

Despite the rapid adoption of LLMs in educational context, there is limited empirical understanding of how these dynamics unfold in authentic, multi-turn human and LLM interactions. Prior work has shown the importance of increasing AI literacy and prompting strategies in shaping user interactions, showing that structured guidance can improve interaction quality and awareness of LLM limitations \cite{koyuturk2025understanding,theophilou2023learning}. In this preliminary study, we focus on whether prompting-style interventions can improve human–AI collaboration in analytical problem-solving tasks and reduce problematic forms of alignment. Through a controlled experiment with real, multi-turn interactions, we examine three aspects pre- and post-intervention: (a) task performance, (b) dependence between user input and AI feedback, and (c) changes in interaction behavior.
Our preliminary results suggest that improving AI literacy and prompting alone may be insufficient, highlighting the need for system-level approaches that better support critical engagement and epistemic independence in AI-assisted decision-making and problem-solving.
\section{Background}
Sycophancy is the tendency of a model to prioritize agreement with the user over accuracy, producing responses that match the user’s views rather than the truth \cite{sharma_towards_2024,richter2025large}. As a result, LLMs may validate user assumptions, mimic errors, or avoid necessary disagreement, particularly in subjective or ambiguous contexts \cite{liu2025truth,sharma_towards_2024}. Recent work shows that sycophancy is not limited to isolated responses but can emerge in interactive, multi-turn settings \cite{liu2025truth}. From a sociocultural perspective, learning benefits from interaction with a partner who offers guidance that is more accurate, reflective, or strategically advanced than the learner’s current understanding \cite{vygotsky1978mind}. LLMs are often assumed to play such a role in educational contexts \cite{koyuturk2023developing}. However, this assumption becomes problematic when model responses stop challenging or extending the learner’s reasoning.
The sycophantic interaction dynamics can lead to epistemic overreliance, where students defer to AI feedback without recognizing its dependence on their own prior responses. Over time, this may disrupt learning by reinforcing incorrect knowledge, disadvantaging less knowledgeable students while benefiting more knowledgeable ones \cite{arvin2025check}. In problem-solving tasks, highly sycophantic systems reinforce user misconceptions and increase reliance on incorrect advice, often without users recognizing the issue \cite{bo2026invisible}. Even when models are capable of identifying correct answers in isolation, they may fail to challenge misconceptions when embedded in real-user-like queries \cite{richter2025large}.
At the same time, user competence and interaction design play a central role in shaping learning outcomes. Novice users often engage with LLMs in an opportunistic manner, struggling to develop systematic and robust prompting strategies, and frequently overgeneralizing from limited successes or failures \cite{zamfirescu2023johnny}. Research on AI literacy and prompting shows that structured guidance can substantially improve interaction quality and empower users to engage more critically and effectively with LLMs \cite{koyuturk2025understanding,theophilou2023learning}.

Although prior work has documented both sycophantic tendencies and cognitive risks, there is limited empirical research examining how these factors emerge in real, multi-turn human–AI collaboration. In particular, it remains unclear how user input shapes AI feedback over time, how this affects decision quality, and whether prompting interventions can meaningfully alter these dynamics.
This preliminary study addresses these gaps by experimentally investigating sycophantic dependence, error propagation in collaboration, and intervention effects in analytical problem-solving tasks, increasing the understanding of how LLMs influence human reasoning and decision-making in practice.
\section{Methods}
We employed a mixed design with a between-subjects manipulation and within-subject task measures. Sixty individuals ($M_{\text{age}} = 50.23$; F:38) with limited experience using generative chatbots were recruited through Prolific from Australia, the USA, the UK, and Ireland, and randomly allocated to either the control group (n = 28) or the experimental group (n = 32). The experiment was conducted in a custom web-based platform built with Django, allowing participants to engage in multi-turn interactions with GPT-4o.

Participants completed four hypothetical survival-ranking tasks designed to assess analytical reasoning and decision-making across different scenarios \cite{yan2025agentic}. Two tasks were completed before the intervention and two after, with task order counterbalanced within each condition. For each task, participants first generated an initial ranking, then collaborated with the AI to discuss and refine their reasoning, and finally submitted their revised and final decisions. These tasks required collaborative judgment under uncertainty, enabling the model to either align with the user or provide more informed rankings. We did not provide GPT-4o with the gold-standard expert rankings to prevent it from emphasizing the correct answers.
During the intervention, all participants first watched a short video introducing general AI literacy and raising awareness of sycophantic tendencies in LLMs. A second video then delivered condition-specific training. The control group received 5 domain-general prompting guidelines, focusing on improving clarity and structure when interacting with LLMs \cite{white2023prompt,koyuturk2025understanding}. Whereas the experimental group received deeper information fostering metacognitive monitoring of AI agreement and user bias, along with 5 sycophancy-specific critical prompting strategies. These strategies emphasized removing personal assumptions from prompts \cite{richter2025large}, explicitly asking for critical evaluation, and requesting supporting evidence \cite{liu2025truth,sharma_towards_2024,doi:10.1126/science.aec8352}.
\section{Analyses and Results}
To quantify performance and AI influence per survival scenario, we computed rank-sensitive agreement scores using Normalized Discounted Cumulative Gain at $k=6$ (NDCG@6), which measures alignment with expert rankings while prioritizing higher-ranked items \cite{wang2013theoretical}.

\begin{figure}[t]
    \centering
    \includegraphics[width=0.9\linewidth]{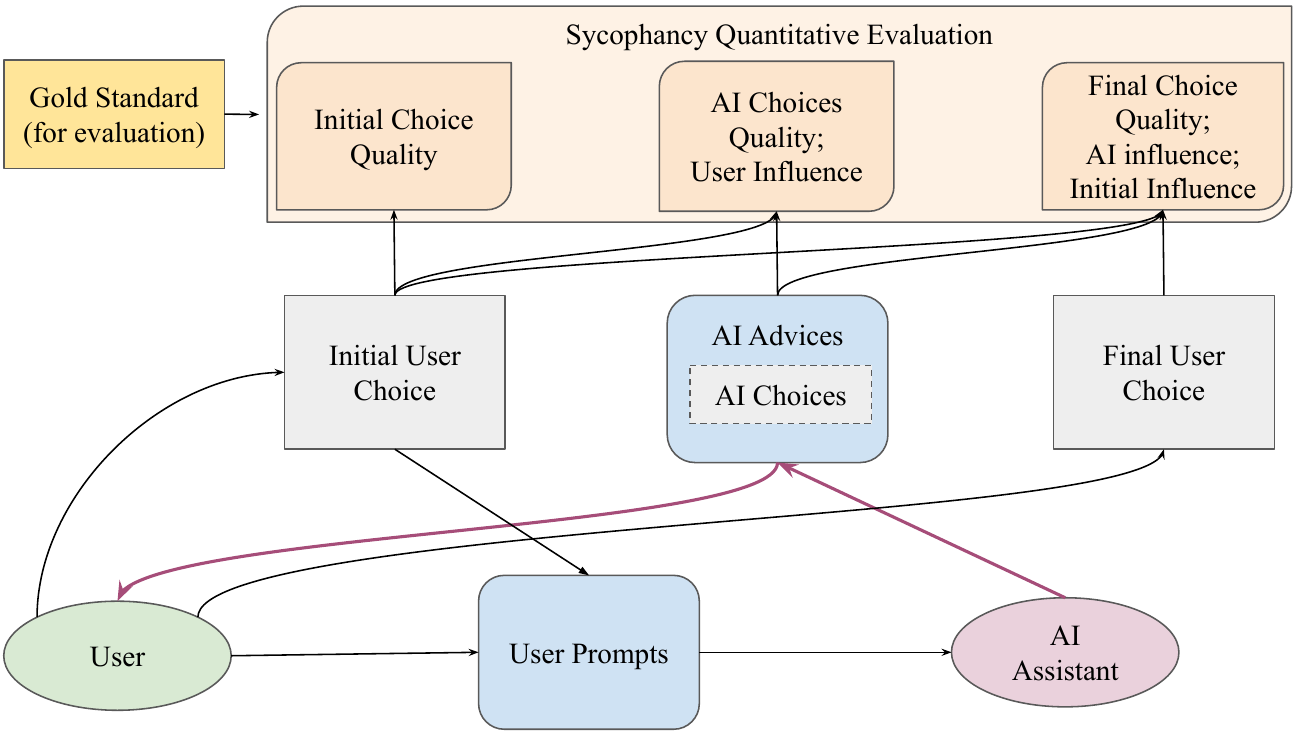}
    \caption{Human–AI interaction loop illustrating potential sycophantic dependence.}
    \label{fig:LBR}
\end{figure}

We derived (a) participants’ \textit{accuracy} as NDCG@6 alignment with expert gold standards (pre- and post-intervention), and (b) \textit{advice quality} as the NDCG@6 of the assistant’s top-6 recommendations, extracted using an LLM-as-judge pipeline \cite{zheng2023judging,ognibene2025use}, in which a separate model (GPT 5.2) analyzed each full conversation transcript and extracted the assistant’s final recommended ranking when it was not explicitly stated in the dialogue (Fig.~\ref{fig:LBR}). A random 10\% of interactions were manually checked to validate the outputs, with no systematic errors observed. 
\subsection{Survival Ranking Performance and Assistance's Advice Quality}
To test whether condition and time (pre- and post-intervention) predicted \textit{final ranking accuracy}, we fit a linear regression predicting \textit{final ranking accuracy} from baseline ranking, condition, time, their interaction, and scenario as a covariate. Users' baseline accuracy significantly predicted final performance, b = 0.414, SE = 0.075, z = 5.53, p < .001, while no other effects were significant (all ps > .44). To assess \textit{AI advice quality}, we regressed it on baseline ranking accuracy and condition. Baseline accuracy significantly predicted advice quality, b = 0.478, SE = 0.180, z = 2.65, p = .008, 95\% CI [0.125, 0.831]. 

\subsection{User-to-AI Dependence and Error Propagation}
We next examined how users’ initial responses shaped assistant behavior. Participants’ initial rankings contained an average of 2.21 non-gold items (SD = 0.93; range = 0–5), indicating frequent suboptimal inputs. The number of non-gold items in users’ initial rankings significantly predicted non-gold items in the assistant’s recommendations, b = 0.264, SE = 0.108, z = 2.44, p = .015, 95\% CI [0.052, 0.476].
To examine how this dependence affected outcomes, we modeled \textit{advice quality} as a function of \textit{user–assistant overlap}, as the proportion of shared items between the user’s initial ranking and the assistant’s advice, and \textit{error carryover}, as the proportion of the user’s non-gold items that were repeated in assistant’s advice. \textit{Advice quality} was positively associated with overall overlap, b = 0.933, SE = 0.128, z = 7.31, p < .001, but negatively associated with the proportion of user errors carried over into the assistant’s advice, b = -0.390, SE = 0.043, z = -9.08, p < .001. 
A similar pattern emerged for final ranking accuracy. Carryover of user errors significantly reduced final performance, b = -0.092, SE = 0.021, z = -4.39, p < .001, whereas overall overlap was not significant.

\subsection{ Effects of the Intervention on Sycophantic Alignment}
We examined whether the assistant’s tendency to align with users’ incorrect inputs changed after the intervention. A binomial model predicting the carryover of user errors showed no significant condition × time interaction, b = 0.152, SE = 0.411, z = 0.37, p = .712, indicating no reduction in general error propagation.

However, using a stricter measure of alignment, whether incorrect items were reproduced at the same rank position, a binomial generalized linear model revealed a significant interaction, b = -1.344, SE = 0.522, z = -2.57, p = .010. Following the intervention, the assistant was substantially less likely to mirror users’ incorrect rankings at the same positions compared to the control condition (OR = 0.26, 95\% CI [0.09, 0.73]). 
A similar pattern emerged for \textit{rank-order alignment}. Spearman correlations between user and assistant rankings of (at least 2) shared non-gold items (n = 87) showed a significant interaction, b = -1.053, SE = 0.314, z = -3.36, p = .001, indicating that the assistant’s ranking of incorrect items became significantly less aligned with the user’s ranking following the intervention. Although the experimental group exhibited higher baseline alignment, b = 0.748, SE = 0.182, z = 4.11, p < .001, the intervention led to a substantial reduction in alignment over time relative to the control condition.

\section{Discussion and Future Work}
This study examined how sycophantic alignment emerges in authentic multi-turn human–AI collaboration and whether prompting-based interventions can mitigate its effects, drawing on a sociocultural perspective on learning as interaction with a more knowledgeable other \cite{vygotsky1978mind}. Our preliminary results showed that LLMs are highly sensitive to user input quality and tend to propagate user errors rather than correct them in real collaboration.
Participants who started with lower-quality rankings received poorer-quality advice, suggesting that the assistant does not operate as an independent corrective scaffold \cite{deng2025deconstructing,vygotsky1978mind}. Instead, it appears to incorporate the user’s initial reasoning, whether correct or flawed, and prioritize the choices that are more frequent or salient in the conversation context, rather than presenting correct but less represented alternatives. The extent to which the assistant carried forward users’ incorrect items strongly predicted both lower advice quality and poorer final performance. This reflects a contextual sycophantic dependence, where the model not only agrees with users but reproduces their errors in ways that degrade decision quality. 

 The intervention did not reduce contextual error propagation. Increasing AI literacy and prompting may not be sufficient on their own to eliminate content-level dependence of the model on users’ initial framing. 
 However, our results showed that the intervention significantly decreased stronger forms of alignment, such as positional mimicry. The assistant continued to incorporate user-provided information but became less likely to mirror incorrect rankings directly. Some apparent sycophantic behavior may stem from general generative biases rather than alignment alone. LLMs may reproduce user inputs because these shift the token distribution, increasing the likelihood of reusing salient context elements without explicit agreement or “pleasing” intent. This aligns with recent work showing repetition and bias can emerge from underlying attention dynamics \cite{huo2026physics}.

As future work, we will analyze the relationship between the adoption of specific prompting strategies and which ones predict improvements, as well as how these strategies evolve with practice and cognitive effort. Overall, this study contributes to the growing literature on human–AI collaboration by providing empirical evidence that sycophantic alignment in real-time, multi-turn interactions leads to error propagation and degraded decision quality, while highlighting the limits of prompting-based interventions in mitigating these effects.

\begin{credits}
\subsubsection{\ackname} This work was supported by IDEAL project, no. 2024-1-IT02-KA220-HED-000251425, AILA project, no. 2025-1-ES01-KA220-HED-000352489, Italian Ministry of University and Research under Grant No. 2023-NAZ-0206, PsyFuture – Dipartimento di Eccellenza 2023-2027, and by Volkswagen Foundation OpenUp Grant Ref. 9E530 Developing an Artificial Social Childhood (ASC).
\end{credits}

%
%
%
\bibliographystyle{splncs04s}
\bibliography{abbrvbib}

\end{document}